\documentclass[10pt,twocolumn]{revtex4}

\usepackage{graphicx,color,comment}

\begin{document}

\thispagestyle{empty}

\title{ \sffamily\Large\bfseries
Structure of Intratumor Heterogeneity: Is Cancer Hedging Its Bets?}

\author{B. Brutovsky}
\affiliation{Department of Biophysics, Faculty of Science, Jesenna 5,
P.~J.~Safarik University, 04154 Kosice, Slovak Republic}

\author{D. Horvath}
\affiliation{Department of Finance, Faculty of Economics, B. Nemcovej 32,
Technical University, 04022 Kosice, Slovak Republic}

\begin{abstract}
Development of resistance limits transferability of most anticancer
therapies into curative treatment and understanding mechanisms beyond
it remains a big challenge.
Many high resolution experimental observations show enormous
intratumor heterogeneity at molecular, genetic and cellular
levels which is made responsible for emerging resistance to therapy.
Therefore, researchers search techniques to influence development
of intratumor heterogeneity, which requires understanding its role within the context
of integrative, logically consistent, framework, such as evolutionary
theory. Although it is agreed that intratumor heterogeneity increases probability
of the emergence of therapy resistant clones, more instructive role
of its structure in the process of cancer dynamics and metastasis
is needed.
In the paper, intratumor heterogeneity is viewed as a product of two,
in general stochastic, processes, evolutionary optimization and
changing environment, respectively. In evolutionary theory, common
risk-diversifying strategy displayed by isogenic populations
in unpredictably changing environments is bet-hedging.
We suggest, that the structure of intratumor heterogeneity
is evolutionary trait evolving to maximize the clonal fitness
in changing (or uncertain) environment
and that its structure corresponds to bet-hedging strategy.
We advocate our view by reviewing and combining important cancer relevant
concepts.
\end{abstract}

\maketitle

\section*{BACKGROUND}

Despite broad acceptance of evolutionary theory as a useful conceptual
framework to understand fundamental features of cancer behavior
\cite{Nowell1976,Crespi2005,Merlo2006}, evolutionary aspect of cancer
is often overlooked in development of novel therapeutic strategies 
\cite{Aktipis2011}.
Although no disqualifying contradiction with evolutionary theory
has been found, the dis-appreciation of therapeutic applicability
of evolutionary theory might come from exaggerated expectations
from too straightforward (or intuitive) applications of basic
evolutionary concepts.
More purpose exploitation of evolutionary nature of carcinogenesis
in therapy necessitates deeper analysis of universal evolutionary
concepts within the context of cancer data.

Many years since postulating the basic principles of evolution by Darwin,
a lot of questions in evolution theory itself have remained open, such as
selection unit, genotype-phenotype mapping, interactions of selection
levels, the role of causation in evolutionary biology, etc
\cite{Noble2008,Nowak2006a,Nanjundiah2011,Okasha2012}.
As biological organisms were, for a long time, the only 'experimental' system
to study evolution, the principles of evolutionary theory merged with
the implementation details. Developing evolutionary theory from the basic
principles through the modern synthesis and genetic determinism towards
the central dogma of
molecular biology, the emphasis of the word 'gene' has, step by step, moved
from its original meaning as 'the cause of an inheritable phenotype characteristic'
to its present meaning as the physical structure. 
This conceptual change is a major source of present confusion in the question
of causation in evolutionary biology \cite{Noble2008,Okasha2012}
and motivates researchers to concentrate on molecular aspects
of cancer evolution.
On the other hand, applications of evolutionary principles outside biology,
in the fields such diverse as optimization, sociology, ecology, etc,
the evolutionary principles must be implemented in the case-dependent way.
Neglecting implementation details, evolutionary theory can potentially predict
universal evolutionary dynamics across wide range of applications 
\cite{Beinhocker2011}.

Many advanced tumors have poor clinical outcome due to development
of resistance to therapy. As experimental studies have revealed enormous
intratumor heterogeneity \cite{Gerlinger2012,Marusyk2012,Yap2012},
it is intuitively agreed that intratumor heterogeneity increases
the probability of harboring a therapy-resistant phenotypes.
Consistently, while normal cells respond very similarly to drugs,
mechanisms of resistance of cancer cells are extremely diverse \cite{Gottesman2002}.
Deeper understanding of intratumor heterogeneity structure and dynamics therefore poses
real challenge to cancer therapy.
Cancer research concentrates mainly on recognizing molecular mechanisms
beyond intratumor heterogeneity and their therapeutic exploitation. More systemic (or functional)
role which intratumor heterogeneity plays in cancer initiation and progression remains less often
studied and relatively poorly understood, limiting usually to its role
in Darwinian evolution of tumors.

Presented genetic data shows that tumors contain complex combinations
of low-frequency mutations thought to drive the cancer phenotypes
\cite{Sjoblom2006}.
Previous studies showed that probably no prototypical cancer genotype
exists and every tumor carries a unique set of mutations, indicating
that multiple genetic pathways may lead to invasive cancer as would be
expected in a stochastic non-linear dynamical system \cite{Gatenby2004}. 
It has been demonstrated that selection for in vitro drug resistance
can result in a complex phenotype with more than one mechanism
of resistance emerging concurrently or sequentially \cite{Hazlehurst1999}.
It was reported that genotoxic stress induces several cell death pathways,
only part of which fall within the classical definition of apoptosis
\cite{Blank2007}.
Accordingly to Witz and Levy-Nissenbaum \cite{Witz2006}, complexity
of the signaling cascades in tumor microenvironment
and the interactive cross-talk between these cascades generate
the feeling that 'anything that can happen – will' and they suggest
to apply tools employed in hyper complex systems analysis \cite{Witz2006}.
Lewis generalizes that, in formally similar system of evolving bacteria,
all of the theoretically logical possibilities of antibiotic
resistance seem to have been realized in nature accomplishing
the same task, which is to prevent the antibiotic from binding
to its target \cite{Lewis2007}.

It was proposed that cancer cells may possibly need only a modest number
of phenotypic traits to deal with all the constraints and evolve
into a tumor \cite{Hanahan2011}. Facing huge
genomic heterogeneity and microenvironmental uncertainty,
the question is if these common "hallmarks of cancer"
are present in all the cells all the time \cite{Floor2012}. Focusing
on a few phenotypic traits would crucially reduce dimensionality
of the relevant search space of all the possible dynamics.
Nevertheless, as any cancer cell can have myriad genetic causes,
as natural selection selects for phenotype, not genotype, and
population changes depend on local environmental selection
forces \cite{Gillies2012}, it leads, at genetic level, to an
undetermined problem. Deeper insight into the problem requires correct
interpretation of intratumor heterogeneity. If interpreted as
a noise hiding a common pattern, the effort to generalize
data from many samples of the relevant cancer type to see the
pattern is justified. If, however, heterogeneity represents
redundancy, i. e. no common pattern exists and each tumor has
unique, nevertheless causative set of genes, to study cancer
by reducing heterogeneity may be a flawed approach \cite{Heng2009}.
If this is the case, the optimization problem solved by cancer
can be viewed as, in a sense, underdetermined, which means that
there are enough degrees of freedom to find multiple reasonable fit
solutions for many environments, i. e. multiple physically realizable
solutions to the relevant mathematical problem.
In \cite{Krakauer2002}, Krakauer and Plotkin applied the quasispecies
model \cite{Eigen1971} and analyzed the evolutionary dynamics of redundancy
and anti-redundancy in general case. They proposed conditions which favor
evolution of redundancy or anti-redundancy, respectively, and list
mechanisms responsible for creating redundancy and anti-redundancy
at the cellular level \cite{Krakauer2002}.

It is intuitively clear that owing to its heterogeneity, evolving population
of cancer cells can absorb and, in conjunction with natural selection, evolve
many alternative solutions to many different environments, representing
an efficient computational system.
Produced by combination of many sources, population heterogeneity becomes
extremely complex statistical quantity.
The measures of heterogeneity may be defined in many different ways most
of them chosen from the viewpoint of the statistical mechanics of non-extensive
systems based on entropy \cite{Buckland2005,Keylock2005,Mendes2008}.
Transforming heterogeneity into a tractable and computable property
of cellular populations provides a rigorous starting point for determining
which variation is random and which is meaningful \cite{Altschuler2010}.

The last decade has witnessed renewed interest of cancer research
community in hierarchical model of cancer, well known as cancer stem cells
(CSC) hypothesis \cite{Reya2001} claiming that cancer cells populations
are hierarchically structured with only a small subpopulation of the cells
able to recapitulate tumor from which they were derived. CSC seems
to possess self-renewal ability and reveal resistance against
conventional therapies \cite{Reya2001,Wicha2006,Clarke2006}.
As soon as the CSC hypothesis was proposed, CSC have become viewed as
the target for therapeutic intervention \cite{Behbod2004,Sugihara2013}.
The strategies rely on the possibility of precise enough splitting
of cancer cells population into CSC fraction and non-CSC fraction. 
The original cancer stem cell model suggested that CSC represent
a subset of cancer cells population which is well distinguishable by
a limited number of cell-surface markers.
During the time, many controversial issues regarding their origin,
proportions in cancer cells population, heterogeneity, flexibility of their
state, etc. have emerged 
\cite{Kelly2007,Shipitsin2007,Adams2008,Shackleton2009,Lathia2011}
and the existence and role of CSC in cancer initiation and progression
remain a topic of intense debate
\cite{Visvader2012,Tang2012,Ni2013,Medema2013,Jordan2009,Clevers2011}.
Accordingly to Badve and Nakshatri \cite{Badve2012}, CSC should be viewed
as representing an aggressive clone that has evolved during tumor progression,
concluding that referring to these cells as CSC is, actually,
a matter of semantics.
Similarly Maenhaut et al. propose that the tumor propagating cells are multiple
evolutionary selected cancer cells with the most competitive properties
maintained by, at least partially, reversible mechanisms, quantitative rather
than qualitative and resulting from a stochastic rather than deterministic
process \cite{Maenhaut2010}. If, however, cancer cell stemness is defined
by function \cite{Jordan2009,Clevers2011,Tang2012,Badve2012} instead
of specific molecular structure, approaches relying 
on the possibility of precise enough splitting of cancer
cells population into CSC fraction and non-CSC fraction may lack efficiency.

There is growing evidence that many solid tumors may be composed of several
distinct subtypes of tumors, which may have distinct CSC \cite{Hope2004,Mather2012}.
Genetic heterogeneity of cancer stem cells \cite{Anderson2011,Notta2011,Shipitsin2007}
provides phenotypic and functional heterogeneity and tempts to designate
the cancer stem cells to be the units of selection in the model of clonal
evolution \cite{Greaves2013}.
From an evolutionary perspective, limitation of self-replicating capacity
to a fraction of tumor cells means that the effective population size
is restricted to this stem-like compartment, rather than encompassing
a bulk of tumor cells \cite{Marusyk2010}.

\section*{REVERSIBLE STATE TRANSITIONS}

A few papers proposed that some isogenic cancer cells
populations consist of phenotypically different subpopulations
and that cancer cells actually switch between different phenotypes
(or cell types) in reversible way
\cite{Chang2008,Quintana2010,Sharma2010,Hoek2010,Chaffer2011,Li2012,Widmer2012},
putting the concept of stemness in question.
In their influential paper, Gupta et al. \cite{Gupta2011} report
observation that the only population of human breast cancer cells
consists of three phenotypically different sub-populations (consisting
of stem, basal and luminal cells, respectively). Studying dynamics
of these cell-types fractions, they found that these stay,
under stationary conditions, in equilibrium proportions.
Moreover, if the cancer cells population is purified for
any of the three cell types, the equilibrium is rapidly re-established
\cite{Gupta2011}.
The progression towards equilibrium proportions would
require implausibly high proliferation rates, therefore they
concluded that the progression towards equilibrium was not due
to differential growth rates of cells in the respective states
but rather to interconversions between cell types \cite{Gupta2011}. 
Summarizing their observations, Gupta et al. proposed a Markov model
of cell-state dynamics \cite{Gupta2011}, which assumes that cells
within a population can exist in any one of $M$ possible stable
cell states (i. e. cell types) and under fixed genetic
and environmental conditions, cells transition from one state
to another with interconversion rates per unit time that
are constant.

Identification of the cell-state dynamics with Markov process
\cite{Gupta2011} enables to study physical and evolutionary aspects
of cancer dynamics separately.
Cancer research concentrates predominantly on the former, physical,
aspects, trying to understand details of molecular mechanisms and
genetic similarities beyond the cell types transitions.
Assuming reported observation, that genetically identical cells under
identical physical conditions differ in their response to a given
chemotherapeutic to an extent that may impact on clinical response
\cite{Saunders2012}, probabilistic nature of Markov model of the
cell-state dynamics may be, eventually, very appropriate.
It follows common mathematical behavior of Markov processes, such
as its convergence towards limiting distribution, which is fully
determined by underlying it transition matrix.
As each element of the transition matrix represents respective
interconversion rate, i. e. the probability of a physical process,
equilibrium cell types proportions in cells population (hence
non-genetic heterogeneity) are fully determined by the transition
probabilities and do not depend on instantaneous cell types
proportions themselves.
In this way, $M$ equilibrium cell types proportions are determined
by $O(M^2)$ interconversion rates which leads, however, to underdetermined
problem for $M>2$.
Despite the fact that physical processes beyond the respective transitions
are certainly not independent each other, huge number of degrees
of freedom of the problem still may leave opportunity to get any
equilibrium distribution of phenotypic proportions by multiple sets
of interconversion rates (i. e. alternative combinations of the respective
processes). Solving  underdetermined inverse problems
is for evolution, in general, an easy task.
Many forms and mechanisms of phenotypic switching have been theoretically studied
\cite{Kussell2005,Kussell2005a,Acar2008,Frankenhuis2011,Libby2011,
Fudenberg2012} and observed at molecular, genetic and expression levels
\cite{Choi2008,Raj2008,Eldar2010,Liberman2011,Pujadas2012}.

\section*{CANCER RELEVANT SELECTION UNIT, TIMESCALE AND CAUSATION}

\begin{figure*}[!t]
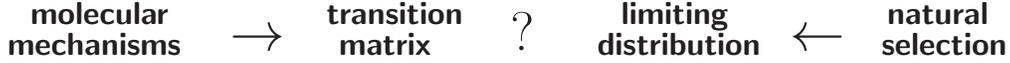

{\sffamily\large\bfseries
molecular\hspace{21mm}transition\hspace{21mm}limiting\hspace{21mm}natural

\hspace{2mm}mechanisms\hspace{21mm}matrix\hspace{22mm}distribution\hspace{16mm}selection
}
\vspace{-6mm}

{\sffamily\huge\bfseries
\hspace{10mm}$\rightarrow$\hspace{30mm}$?$\hspace{34mm}$\leftarrow$
}

\vspace{2mm}

\caption[]{\noindent\small 
Question of causality: is limiting distribution of the state-cell dynamics
\cite{Gupta2011} just 'lucky' coincidence rooted incidentally in the complexity
of epigenetic landscape or, on the contrary, is the transition matrix 
evolutionary outcome of specific tumor microenvironment?
}
\label{Causation}
\end{figure*}

Despite often referred observations that cancer cells actually switch
between different cell types in reversible way 
\cite{Chang2008,Quintana2010,Sharma2010,Hoek2010,Chaffer2011,Li2012,Widmer2012},
heterogeneity is, in cancer biology, traditionally attributed
to genetic variance, implicitly implying one-to-one genotype-phenotype
mapping. Many authors propose that genetic heterogeneity
is unlikely to be the major contributor to phenotypic heterogeneity
in general, but, underlying heritable differences, it fuels tumor
evolution \cite{Marusyk2012}.
As non-genetic mechanisms, such as gene expression noise and multiplicity
of stable states in gene networks, are responsible for phenotypic
identities of normal cells, it was suggested that non-genetic heterogeneity,
can contribute to somatic evolution of cancer cells, hence accelerating tumor 
progression and development
of therapeutic resistance \cite{Brock2009,Huang2009_Development,Huang2012}.
Recently, the difference between two types of cancer cell instability,
genetic and non-genetic, was accented and hierarchical link between
the corresponding spaces, the fitness and epigenetic landscapes,
was proposed \cite{Huang2013}. Therein, each point in the fitness landscape
(i.~e.~genome) provides epigenetic landscape of unique topology.
Due to its mathematical complexity, each epigenetic
landscape contains, stable areas (attractors) around stable
cell-states \cite{Huang2013}, viewed as cell types or phenotypes (disputed
below).
Non-occupied attractors are not exposed
to selection and, consequently, are not evolutionarily harmonized with the needs of
the tissue \cite{Huang2013} and stay pathological (or cancerous).
If epigenetic landscape, due to genetic mutations or tumor microenvironment,
changes, probability of the cell finding itself in cancerous attractors
may increase \cite{Huang2013}.
Considering the timescale in which mutations spread in a cell population,
non-genetic instability is made responsible for heterogeneity
of cancer cell populations \cite{Huang2013}.
This is consistent with the observation that the population of isogenic
cancer cells purified for one of the stable cell-states re-establish
equilibrium proportions of the cell types too fast to be explained
by differential growth alone \cite{Gupta2011}.
Consequently, the role of somatic evolution in cancer progression is
put in doubt \cite{Huang2013}.

Keeping in mind Theodosius Dobzhansky's statement
{'Nothing in biology makes sense except in the light of evolution',
we propose the eventual role played by non-genetic heterogeneity
in somatic evolution of cancer.
To understand intratumor heterogeneity from evolutionary viewpoint,
fundamental aspects of natural selection, namely selection unit,
timescale and causation must be reconsidered.
If somatic evolution is applied to explain cancer, cancerous features
are implicitly attributed to a cell which is taken as cancer-relevant selection
unit and its fitness is identified with its reproduction capability.
Assuming one-to-one genotype-phenotype mapping, current cancer research focuses
on differences in metabolic pathways responsible for cancer phenotype of the
cell in order to distinguish cancer and normal cells as well as to predict
proximate behavior of cancer cells.
However, the focus on differences between the normal and cancer cells
may be not adequate enough, as it was observed that (at least) some isogenic cancer
cells populations consist of phenotypically different subpopulations \cite{Gupta2011},
which implies cell's multistability (or one-to-many genotype-phenotype
mapping).

The term 'phenotype' is often used in very intuitive way, denoting
usually observable traits of the cell, which are assumed to be static.
If alternative sets of observable traits expressed by the same
genome is observed, each of them is viewed as an alternative phenotype,
and the change of phenotype without genetic cause is interpreted as
'phenotypic switch'.
Despite being appropriate in many biological contexts, if evolutionary  
theory is to be applied in cancer biology, the term 'phenotype' must 
be reconsidered. Evolution is based on the general
premise that any population of entities which reveal variation,
reproduction, and heritability may evolve \cite{Lewontin1970}.
The original biological meaning of a gene referred to the cause of an 
inheritable phenotypic characteristic \cite{Noble2008}. It implies, that
the genome, as a set of genes, refers to the cause of inheritable phenotype
as a whole. The above evolutionary premise is formally fulfilled not only
when the unique cell-type expressed by the genome is taken as phenotype,
but, as well, if the epigenetic landscape, corresponding to the genome
as its unique cause, is assumed to be the phenotype.
In the latter context, the term 'phenotype' encompasses all the relevant
features of the epigenetic landscapes, such as the repertoire of attractors
(i. e. cell types), the heights of barriers between them, etc.

The fitness of the genome relates to the phenotype-relevant timescale.
At proximate timescale, its phenotype (i. e. a cell-type) is the  
outcome of specific molecular mechanisms and its fitness is obvious.
At longer timescale, after the genome produced a lineage of reasonable size,
the size of its clone corresponds to the fitness of the genome (at this
timescale).
Obviously, evolutionary success (or failure) of the cell at short timescale
does not necessarily correlate with its evolutionary
success on long timescale.
To quantify possible outcome of lineage (or clone) evolution
in a more quantitative way, Palmer and Feldman introduced two metrics,
$k$-fitness and $k$-survivability \cite{Palmer2012}.
The former quantifies probability of increase of the size of the
respective lineage after $k$ generations, the latter relates to
the likelihood that the species will avoid extinction after $k$ generations.
If $k$ increases, $k$-fitness of the cell depends more and more
on the eventual interaction of the cells in the clone.
Regarding the timescale at which cancer has effects (cancer-relevant
timescale), the clone seems to be more relevant structure to determine
the genome fitness.

Leaving precise quantification of cancer-relevant timescale to further
research, we ask how general evolutionary top-down causation \cite{Okasha2012}
applies in cancer?
At proximate timescale, the genomes' fitnesses depend on their respective
'built-in' molecular machinery.
On the other hand, molecular machinery has been selected accordingly
to the evolutionary advantage they conferred to the ancestors of the
respective genomes (Fig.~\ref{Causation}) in the past.
The above reconsideration of cancer-relevant selection unit, timescale
and causation imply what is the selection force evolving epigenetic landscape
and what is the optimum repertoire of stable cell-states (i. e. cell types)
conferring to the cell the highest fitness at cancer-relevant timescale.

\section*{BET-HEDGING}

Markov model of the cell-state dynamics by Gupta et al. \cite{Gupta2011}
implies that non-genetic heterogeneity in isogenic cancer cells population
(i.~e.~cancer clone) is actually determined by the probabilities of transitions
between different
cell types, i. e. by the transition matrix of Markov process, and are not
bound to genetic polymorphism. The cell types correspond to stable
cell-states (or attractors) in the epigenetic landscape as conceptualized
by Huang \cite{Huang2013}.
In this way, genetically coded transition matrix prescribes the sizes
of the cell types fractions (hence non-genetic heterogeneity) in any possible
tumor microenvironment, and, from the evolutionary viewpoint, the fitness
of the genome at the cancer relevant scale. As the cell types, in general,
differ in their growth (and other) properties, the distribution of the
cell states in the clone (non-genetic heterogeneity) becomes
evolutionary important.
To sum up, the cells evolve transition matrix producing the proportions
of the cell types so that the clone increases at maximum rate. During
their evolution the cells inevitably interact, which is traditionally
viewed as 'cooperation' in the case of normal cells, while the cancer cells
are often said to behave 'selfishly',
\cite{Greaves2007} revealing specific prototypical features \cite{Hanahan2011}.
It was, however, proposed as well, that genetically
distinct tumor cells cooperate as well to overcome certain host defenses
by exchanging different diffusible products \cite{Axelrod2006}.
Cooperation does not imply equality of the cells in the clone but
rather a specialization of the cells to diverse roles,
which increases clonal fitness. In the conceptual model of group
selection \cite{Nowak2006a}, pure cooperator groups grow faster than
pure defectors groups, whereas in any mixed group, defectors reproduce
faster than cooperators, i. e. selection on the lower level (within
groups) favors defectors, whereas selection on the higher level
(between groups) favors groups consisting of cooperators
\cite{Nowak2006a}.

Inspired by the above model of cooperation \cite{Nowak2006a}, we suggest
that the cancer cell, its clone being the evolutionary winner at the relevant
timescale, produces the clone of well 'cooperating' cells, and propose what
kind of cooperation between cancer cells can be expected.
Instead of interpreting specific biochemical reactions as cooperative
or selfish, cooperation is viewed in a more general way - as a coordinated
action. This view enables its statistical interpretation as mutually
correlated dispersal of the cell states in the state space.
The observation by Gupta \cite{Gupta2011} implies, that the cell types
equilibrium (i. e. non-genetic heterogeneity) in isogenic clone
of cells can be established by switching between a few possible states
in a reversible way.
Assuming that the cells in different states differ in their proliferation
efficiency, intraclonal cooperation is determined by the cell-states
heterogeneity when the cells in different states contribute to the clone
growth by, in general, different amounts.
Different environments confer different growth to the clones with different
heterogeneity structure - in some environments clones consisting of the cells
in the same state may provide faster clonal growth, while in the others
more specialization (bound to specific structure of cell-states heterogeneity)
leads to faster growth of the clone.
In this view, non-genetic heterogeneity becomes evolutionary trait and
the fundamental question arises what heterogeneity structure gives
to the cell the highest fitness (measured as the size of its clone).

From the viewpoint of evolutionary biology, phenotypic heterogeneity
is an adaptation to environmental uncertainty and phenotypic diversification
enables species to survive environmental adversity \cite{Fraser2009}. 
Each phenotype (or, in here applied conceptualization, cell-type)
proportion can be viewed as an investment of certain portion of population's
reproductive effort \cite{Donaldson2008}.
Two fundamental evolutionary strategies of population adapting to environmental
uncertainty are well known \cite{Donaldson2008}. The former, generalist,
produces constant phenotype which has been reasonably fit in any relevant environment.
The latter, 
the bet-hedging strategy, generates non-genetic phenotypic diversity
in the population producing phenotypes accordingly to probability distribution
matching the distribution of the environments rewarding (in a sense of
fitness) the respective phenotypes in the past.
One can find instructive analogy with risk diversifying strategy
in portfolio management. Facing uncertain future, investor divides
his budget into a few (or many) assets instead of the only, whatever
probable, asset to protect himself
against fatal loss. Optimum investment strategy must somehow balance
predictability of assets' prices with the cost of portfolio's restructuring.
Optimal portfolio depends on the dynamics of the trends, and, at the same
time, the investor's capability to restructure his 'portfolio'.
It is of the utmost importance that the bet-hedging strategy is realized
as alternative expressions of the only selection unit, not as a form
of genetic polymorphism \cite{Seger1987}.

In laboratory studies of yeast and bacteria, the rate of phenotypic switching
has appeared to adjust to match the frequency of environmental changes
\cite{Liberman2011}.
In their model of survival in changing environments \cite{Kussell2005a},
Kussell et al. demonstrated that the optimal switching between normal cells
and bacterial persister cells, characterized by slow growth and increased
ability to survive antibiotic treatment, depends strongly on the frequency
of environmental change and only weakly on the selective pressures
of any given experiment \cite{Kussell2005a}.
It is consistent with the finding that a critical feature of the process
of tumor progression is selection of cells that can escape from resource
limitations by achieving a relative microenvironmental independence
\cite{Anderson2009}.

To sum up, the bet-hedging strategy \cite{DeJong2011} is a universal
risk-diversification strategy evolved in the populations which face
uncertain future and/or environment \cite{Crean2009,Forbes2009,Beaumont2009}.
Assuming formal similarity of evolving cancer cells population with
the above evolutionary systems, 
we identify non-genetic heterogeneity corresponding to the observed equilibrium
distribution of the cell types in isogenic cancer cells population
\cite{Gupta2011} with the bet-hedging strategy. This identification
is motivated by the observation that a rapid progression towards
equilibrium proportions would require implausibly high proliferation
rates and cannot be explained by differential growth rates of the
clones of cells in the respective states, but rather to interconversion
between states \cite{Gupta2011}. Moreover, non-genetic heterogeneity
develops at the timescale in which mutations barely spread in a cell
populations \cite{Huang2013}.

To propose the role of non-genetic heterogeneity in clonal evolution
of cancer, selection force which pushes the cell types proportions
into the "optimum investment profile" in the respective environment, corresponding
to the respective environmental frequencies, must be identified.
Applying the above conceptualization by Huang \cite{Huang2013}, each
the genome represents epigenetic landscape of unique topology, with
the repertoire of attractors (i. e. cell types) as a fundamental
mathematical feature. These attractors are separated by the barriers
which heights determine probabilities of transition between them.
In the conceptualization by Markov model \cite{Gupta2011}, the repertoire
of attractors corresponds to the limiting distribution of Markov model
and the heights of the barriers to its transition probabilities.

On the other hand, environment predetermines optimum investment profile,
which, if occupied by the respective (optimum) proportions of cancer cells,
maximizes growth of the clone.
In bet-hedging theory, the strength of selection towards
the optimal bet-hedging strategy depends on how far the residents
are from the optimal investment profile \cite{Donaldson2008}.
Straightforwardly, the fitness of the cancer cell is given by
the deviations of the cell types proportions in its clone
from the optimum bet-hedging proportions at cancer-relevant
timescale.
As the cell types proportions (hence the clone's cell-state heterogeneity)
are determined by the rates of interconversions between states
\cite{Gupta2011}, i. e. by the probabilities of specific physical
processes, the genetically coded molecular mechanisms are under
selection pressure preferring those which provide heights of
the barriers between attractors leading to the optimum cell types
fractions as required by the environment dynamics.

\section*{EVOLUTIONARY OPTIMIZATION} 

\begin{figure*}[!t]
\includegraphics[scale=0.44]{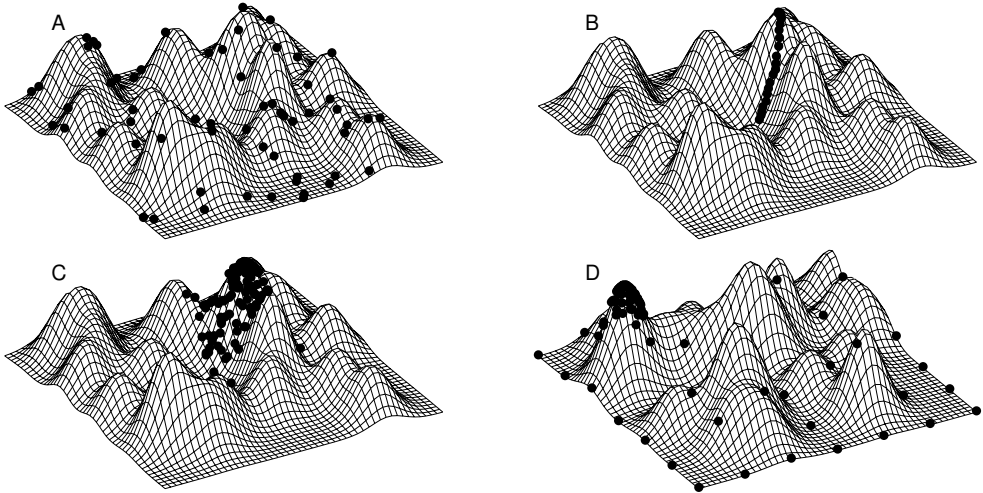}
\caption[]{\noindent\small 
Examples of different sampling strategies. (A) blind search which maximizes
exploration but completely omit exploitation, (B) gradient search maximizing
exploitation with exploration represented just by the
determination of gradient, (C) simulated annealing \cite{Kirkpatrick1} trying 
to balance exploration with exploitation. In (D), constant 'sentinels' are
added to prevent shrinking of the search space \cite{Morrison2004}.
In A,B and C, the sampling points
are generated in serial way, in D they represent population and are
evaluated in parallel.}
\label{Sampling_Strategies}
\end{figure*}

In 1932, Sewall Wright conceptualized evolution as a search process through
the astronomically huge search space of all the possible combinations
of genes \cite{Wright1932}. Assigning a fitness value (a quality
measure) to each genetic combination, the evolution was identified
with a search for the highest peak (maximum fitness) in so-called fitness
landscape. In this way, Wright linked
evolution with the field of optimization as conceived in engineering
and economy \cite{Beinhocker2011}, initiating the new branch
of stochastic optimization techniques, presently known as evolutionary
algorithms (EA) \cite{Holland1975}.
Despite the fact that in evolutionary optimization one purposely 
applies evolutionary principles to evolve population of candidate
solutions, while biological evolution is straightforward consequence
of mere existence of the population of biological replicators (genomes),
both the processes depend on universal aspects of fitness landscape.
EA analyze the above features in implicitly abstract way, while in biological
evolution they are bound to specific molecular machinery which complicates
their analysis.
Being applied in many different contexts, EA have significantly enriched
evolutionary theory by sharpening the above "substrate-free" aspects
of evolutionary dynamics.

In real-world optimization problems (including biological evolution)
each fitness evaluation requires nonzero resources and must be attained
within affordable time interval, which limits number of fitness evaluations.
On the other hand, keeping in mind the fitness landscape uncertainty, rational strategy
is exploring, with some probability, also not yet evaluated parts
of the search space. Uncertainty of stationary fitness landscapes
decreases proportionally to the number of fitness evaluations, therefore
the typical strategy is to allocate, during the optimization, increasing
number of trials to the observed best solution, and let explorative
power of the algorithm vanish. 
But the question emerges how fast should the ratio between 
exploration and exploitation aspects decrease during optimization?
The problem is known as the exploration vs. exploitation dilemma
(or the optimal trial allocation) and optimization techniques differ
in their way of solving it (Fig.~\ref{Sampling_Strategies}).
Blind search (Fig.~\ref{Sampling_Strategies}A) resigns exploitation
preventing it from sticking in local optima but, at the same time,
results in very low efficiency. Gradient search
(Fig.~\ref{Sampling_Strategies}B) always continues uphill, maximizing
exploitation but it fails with
high probability if the fitness landscape contains more local optima.
Simulated annealing \cite{Kirkpatrick1} (Fig.~\ref{Sampling_Strategies}C)
represents some compromise, starting as random search, decreasing continuously
probability of acceptance of less fit points, converging, eventually,
in global optimum.
Obviously, efficiency of sampling strategies crucially depends
on dimensionality, ruggedness, modality, stationarity, etc
of the respective fitness landscape.

The problem of the optimal allocation of trials is especially
challenging if the fitness landscape is changing, which is the crucial
feature of biological fitness landscapes. Facing uncertain future,
optimization procedure must maintain nonzero explorative ability.
To be efficient in changing fitness landscape, an optimization procedure
must i) detect change in the fitness landscape (exploration),
and, ii) appropriately respond to it (exploitation). Due to nonzero
detection and response time, tracing the optimum in dynamic optimization
problem always expects nonzero time correlation of fitness landscape
which can be exploited. 
The maximum entropy principle states, that, if the probability distribution
of random variable is not known, the probability distribution which best
represents the current state of knowledge is the one with the largest
information theoretical entropy. It implies preemptive distribution
of the trials in evolutionary algorithms designed for dynamic environments.
They keep fraction
of candidate solutions, denoted as sentinels, unchanged \cite{Morrison2004}.
The sentinels are population members that are statistically
reasonably (in the above sense) distributed through the search space upon
initialization, and kept in population to produce new population members
through selection, but themselves are neither mutated nor
replaced \cite{Morrison2004}.
The sentinels themselves are, in average, neither more nor less fit than
other solutions in the population. Their added value consists in preventing,
in conjunction with the other sentinels, shrinking of the search space
(Fig.~\ref{Sampling_Strategies}D), i. e. maintaining the exploratory power
of the algorithm.

\section*{SUMMARY}

Clonal evolution model for tumor progression by Nowell \cite{Nowell1976}
says that within a population of tumor cells natural selection, which
favors cells that have acquired the most aggressive phenotype, occurs.
To make this predication more applicable, better understanding what
'the most aggressive' means is necessary. As soon as the 
CSC hypothesis was proposed \cite{Reya2001}, the subpopulation
of CSC was alleged to be the most competitive tumor fraction,
and, consequently, tumor evolution was attributed to it.
Despite its universality is still debated, the CSC hypothesis
implies basic hierarchical structure of intratumor heterogeneity.
Evolving cancer cells population
reveals the structure of heterogeneity (CSC vs. non-CSC fractions)
similar to the structure of population heterogeneity purposely
applied in evolutionary optimization to make optimization
procedure more efficient in changing fitness landscapes
\cite{Branke2002,Morrison2004}. This is not surprising,
as the link between evolution and optimization was made long
time ago \cite{Wright1932}.
To sum up, the intratumor heterogeneity structure of cancer
clone (being the evolutionary winner) has balanced optimally
exploration of the search space with its exploitation during
cancer-relevant time period.

Cancer heterogeneity has been for a long time viewed as the consequence
of genetic heterogeneity. Presently, an opinion is emerging that
the stemness of cancer cells corresponds more to function state
instead of distinguishable genetic (or epigenetic) pattern.
This view is supported by reversible stochastic switching between
cell states in cancer cells population
\cite{Gupta2011,Chang2008,Quintana2010,Sharma2010,Hoek2010,Chaffer2011}.
The results by Gupta et al \cite{Gupta2011} show, that the specific cancer
genome encodes, at the same time, three alternative stable cell types,
one of them being the 'stem' state, and switching between them creates,
in stationary environment, equilibrium proportions of the cell types.
In this way non-genetic heterogeneity is responsible for cancer and
some papers present doubts about the role of somatic evolution
in cancer.

In the paper, the equilibrium cell types proportions are viewed to be
more refined structure of intratumor heterogeneity, beyond the basic
division into CSC and non CSC fractions. The genome stays the main
protagonist (i. e. selection unit) in the evolution of cancer cells,
with non-genetic heterogeneity of its eventual clone being the
crucial adaptive trait at cancer-relevant timescale.
From the viewpoint of evolutionary biology, phenotypic heterogeneity
is an adaptation to environmental uncertainty and phenotypic
diversification enables species to survive environmental adversity.
One of the observed and well studied strategies of population
diversification in changing environment is the bet-hedging strategy
\cite{Seger1987,DeJong2011,Donaldson2008}, which divides reproductive
investment in each environment to fit the respective environmental
frequencies in the past.
Straightforwardly, we suggest the hypothesis that non-genetic heterogeneity
in cancer cells population evolves towards the universal bet-hedging
diversification strategy. Affirmative answer
to the hypothesis necessitates determination of the optimum bet-hedging
profile in specific cancer case during some relevant past, which
is the task far from trivial. 
Bound to the same evolving structure, the genome, the two components
of phenotypic heterogeneity (genetic and non-genetic) interact in
complex way, which dramatically complicates their more rigorous analysis.

How can be the affirmative answer to the question in the title helpful?
Having attributed cancer dynamics to an appropriate universal dynamics
(such as here proposed bet-hedging), one can, eventually, apply its general
features to influence the process by modifying fitness landscapes in
mathematically more purposeful way. It is known that diversification
strategy adopted by evolving populations depends on the environment
\cite{Donaldson2008}. Under some environments
dynamics the generalist strategy are more successful than bet-hedging
\cite{Donaldson2008}. Straightforwardly, purposeful manipulation
with statistical features of environment (hence fitness landscape)
dynamics, may provide diversification strategy
which is less fatal and/or better controllable.

\section*{ACKNOWLEDGMENTS}

This work was supported by the (i) Scientific Grant Agency of the Ministry of Education
of Slovak Republic under the grant VEGA No. 1/0370/12, (ii) Agency of the Ministry
of Education of Slovak Republic for the Structural funds of the European Union, Operational
program Research and Development (SEPO II ITMS code: 26220120039),
and (iii) Slovak Research and Development Agency under the contract APVV-0242-11.



\end{document}